# Corner-Sharing Tetrahedra for Modeling Micro-structure

Meera Sitharam, Jeremy Youngquist, Maxwell Nolan, Jörg Peters

*University of Florida*

**Abstract**

State-of-the-art representations of volumetric multi-scale shape and structure can be classified into three broad categories: continuous, continuous-from-discrete, and discrete representations. We propose modeling micro-structure with a class of discrete Corner-Sharing Tetrahedra (CoSTs). CoSTs can represent bar-joint, tensegrity, line-incidence, and similar constraint systems that capture local physical constraints and global multi-scale properties for design and analysis. The paper develops a palette of simple geometry processing operations on CoSTs including graph manipulation, hierarchical refinement, randomization, and generating associated continuous representations.

## 1. Introduction and Motivation

Advances in composite materials and additive manufacturing permit, in principle, precise generation of micro-structure and shape at multiple scales. This requires however new least-cost geometry representations capable of dealing not just with surfaces or partitions of the macro-geometry, but with massive internal micro-structures. And micro-structure requires corresponding geometry processing tools to manipulate such representations to meet interior and boundary physics and macroscopic shape requirements. Current micro-structures fall into three broad categories: *continuous, continuous-from-discrete, and discrete representations* (see Section 1.2 below). Arguably the simplest representation of micro-structure is a geometrically realized graph of distance relations (edges) between geometric primitives (nodes). The advantage of such a minimal and abstract representation is ability to model many meta-material properties and behavior at multiple scales purely as properties of the underlying graph, i.e. independent of geometry of geometric primitives. Since materials should neither be over-constrained (internal stress) nor under-constrained (mechanisms), we focus on well-constrained graphs. To avoid the difficult realization problem, we initialize material as a regular Kagome partition and prove that simple triangle and tetrahedral flips generate a rich class of models. Kagome structures have been extensively studied by experimentalists [13, 51, 49] due to their desirable multi-physics properties. While many other meta-material truss-based architectures do not preserve joints from previous levels in a hierarchy, so complicating theoretical predictions of physical properties [28, 48, 45], Kagome lattices have a natural refinement that preserves the coarser structure.

### 1.1. Contributions

- We introduce geometry processing of *Corner-sharing Triangle and Tetrahedra (CoST)* whose graphs have baked-in desirable physical properties.

- We create a *complete palette of CoSTs* starting with Kagome structures. In the bivariate case this is as rich as all triangulations of a given point set.

- We provide *hierarchical CoSTs* by constructive refinement that preserves basic properties and neighbor relations.

- We show how to fully *explore the CoST design space* by *local* graph and geometry operations. (This also yields a process for modeling irregular micro-structures such as particle and fiber matrix composites using CoSTs.)

- We provide efficient graph and geometric operations for consistently *joining CoSTs*.

- We pair CoSTs with *continuous, spline-based representations, both implicit and parametric*.

### 1.2. Micro-structure modeling

Micro-structure representations fall into three broad categories: continuous, continuous-from-discrete, and discrete.

*Continuous* functions and fields allow for many approaches to match boundary specifications, e.g. conforming, meshless, etc.. Continuous representations are prevalent in the area of topology optimization. Combining a nonlinear objective function with linearized constraints, topology optimization can link shape and micro-structure at multiple scales and is coupled with finite element physics computations to incorporate physics and macroscopic shape, see e.g. [3, 5, 6]. However continuous representations do not easily encode or reveal discrete connectivity, such as provided by joining beams and struts. Enforcing consistent material connectivity typically requires nonlinear or integer constraints.

To represent random micro-structure, *continuous-from-discrete* representations first generate pre-micro-structures via Poisson field smoothing and Voronoi subdivision, based on a discrete set of positions of (ellipsoidal) shapes [21, 12]. These pre-micro-structures are subjected to (simulations of) sintering and other processes to yield *effective micro-structures*. [47, 12]. Tailoring the micro-structures of 'matrix composites' to physics and macroscopic shape then involves choosing controllable particle and matrix parameter values and a stochastic model of distribution over *pre-micro-structures*. Due to limited control over parameters and the need for a simulation pipeline, this approach is generally used only for effective bulk properties of micro-structure [22].

Our focus is on a third class, *discrete* representations (D-reps). D-reps have been used by physicists in the meta-material design community [38, 37] to model shape or physical micro-structure features as geometric primitives, and assert a graph or network of local, pairwise constraints among them [11]. Geometric primitives and constraints can so encode not only spatial but also mechanical, electromagnetic and thermal local interactions (at equilibrium). D-reps can capture interactions between particles in a composite material [9, 10, 17]. For most types of bivariate D-reps and certain classes of trivariate D-reps, these properties can be measured using efficient combinatorial algorithms. Yet, compared to continuous and discrete-continuous representations, systematic D-rep design tools are few and adhoc.

### 1.3. Organization

Section 2 lays out notation and summarizes existing theory. Section 3 outlines required design operations. Section 4 explains the operations. Section 5 explains show that surfaces and volumetric splines can be associated with the proposed representation. *All figures except Fig. 1d were generated with components of our prototype microstructure design and modeling environment.*

## 2. Theory of discrete constraint-based representations

A *geometric constraint system* (GCS) relates *geometric primitives* (e.g., points, lines, planes, rigid bodies, conics) [16, 15, 2] as variables. The setting can be Euclidean or non-Euclidean geometry of fixed dimension, and constraints specify geometric pairwise relationships among the primitives: they can be binary (e.g., incidence, perpendicularity, tangency) or metric (e.g., distance, angle, orientation) and may be either equalities or inequalities. Typically, a constraint is expressed as a set of quadratic polynomials in the geometric primitives. The combinatorics of a GCS are expressed in a *constraint graph* where each vertex represents a geometric primitive and each edge represents a constraint on a pair of primitives.

A *realization* of a GCS is a placement (or configuration) of the geometric primitives that satisfies the constraints. A realization of a GCS can be found algebraically by solving the system of equations corresponding to the GCS, where the variables are the coordinates of the geometric primitives. Generally, the realization space is taken modulo some group of *trivial motions* such as translation and rotation.

A *D-rep* is a realization, together with the underlying geometric constraint graph.

### 2.1. Types of D-reps for representing micro-structure

The most common GCS is the *Euclidean distance, or bar-joint constraint* system. The geometric primitives are points (called "joints"), the constraints are specified distances between points (called "bars") and the ambient space is $\mathbb{R}^d$. The constraint graph $G = (V, E)$ associates a vertex to each joint and an edge $(u, v)$ to each bar constraining the joints represented by vertices $u$ and $v$. Then the bar-and-joint constraint system of $G$ can be defined as a tuple $(G, \delta)$ where $\delta : E \to \mathbb{R}$ assigns distance values to the bars. A configuration of the joints given by a map $p : V \to \mathbb{R}^d$ and is a bar-joint D-rep realization of $(G, \delta)$ if the distance between $p(u)$ and $p(v)$ is $\delta(u, v)$ for all $(u, v) \in E$.

*Tensegrity* D-reps generalize bar-joint D-reps in that the underlying GCS involves inequality as opposed to equality constraints. Some edges of the constraint graph, called *struts*, have distance lower bounds and others, called *ties* have distance upper bounds. For geometric primitives that are rigid *bodies* in $d$ dimensional Euclidean space a *pin* constraint can be placed between two bodies by picking a point on each and making them incident. Such a system describes a *body-pin constraint system*. Realizations in this case are solutions to the system of incidence equations, i.e., placements of the bodies (transformations in the special Euclidean group $SE(d)$) that satisfy the pin incidences. *Body-bar* D-reps replace the pins by bars. It turns out that body-pin D-reps can be converted to equivalent bar-joint D-reps or a body-bar D-rep. See Figure 1 for examples. A *pinned line-incidence system* $(G, \delta)$ is a graph $G = (V, E)$ together with parameters $\delta : E \to \mathbb{R}^2$ specifying $|E|$ *pins* with fixed positions in $\mathbb{R}^2$, such that each edge is constrained to lie on a line passing through the corresponding pin. A D-rep in this case is a set of pinned lines that are coincident whenever the corresponding edges share a vertex. See Figure 1c for an example constraint graph and D-rep realization for a pinned line-incidence D-rep. The *polar dual* D-rep treats the vertices as lines, the edges as incidence constraints between pairs of lines and the pins as linear *sliders* on which the points of incidence are constrained to lie. There is a bijection between pinned line-incidence D-reps and their polar duals. We list all these examples to illustrate that *different types of D-reps can have the same underlying geometric constraint graph!*

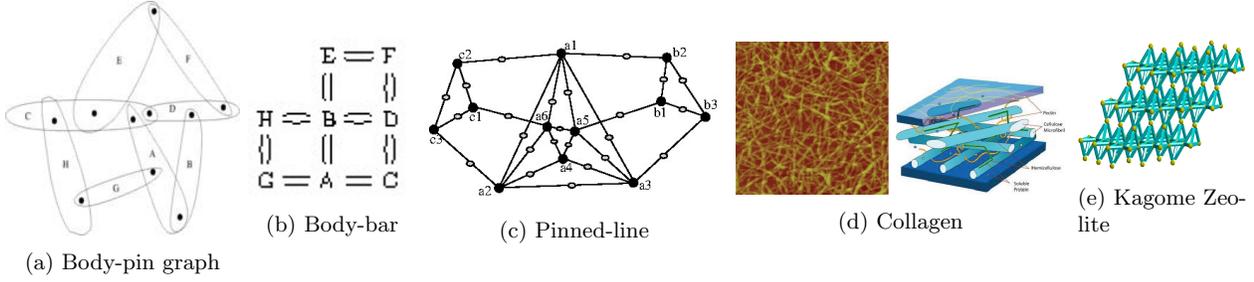

(a) Body-pin graph  (b) Body-bar  (c) Pinned-line  (d) Collagen  (e) Kagome Zeolite

Figure 1: Examples of D-reps. (b) is the body-bar representation of (a). Collagen (d) can be modeled as a pinned-line incidence. (e) shows the bar-joint representation of quartz-like materials.

### 2.2. Physical Realizations of D-reps

The GCS names bar-joint, body-pin, etc. hint that D-reps can be endowed with a physical or material realization such as pin-jointed (universal) body or truss structures (see Figures 1a, 1b). Tensegrity D-reps containing only struts can, for example, represent packed incompressible disks or spheres. If the disk/spheres are sticky, then a bar-joint D-rep suffices. Pinned-line incidence systems [2, 43] cellulose). Cross-links force fibers to touch and run through constraining holes at the pin positions, but leave them free to slide and rotate. Fig. 1d illustrates some physical realizations. Beyond mechanical constraints, geometric constraints can be used to represent energy barriers [42].

### 2.3. Abstract Rigidity

There is close relationship between properties of D-reps and static physical properties of their material realizations. If the number of constraints in the GCS matches the number of variables representing the geometric primitives, modulo the trivial motion group, and all constraints are *independent* then the D-rep is *minimally rigid* (a.k.a. isostatic or well-constrained.) Any deficit in the number of independent constraints yields *flexible* realizations [1] (Note that a D-rep with dependent constraints can simultaneously be flexible and overconstrained). For $d = 2$, bar-joint, tensegrity, body-pin (converted to bar-joint), direction line-incidence, point-line distance-angle and other D-reps [11, 40, 36, 19, 9, 14] share a purely combinatorial and algorithmically efficient characterization of generic rigidity. A famous special case of such a characterization for $d = 2$ based on tight bounds on sparsity of the constraint graph is by Hilda Geiringer [39] for bar-joint D-rep and is called Laman's theorem [24]). When such a characterization exists, the corresponding D-reps and their underlying graphs are called *nice*. For many nice D-reps, Theorem 1 generalizes beyond the Euclidean plane when the geometric primitives and allowable motions lie on a sufficiently smooth manifold. For trivariate D-rep, Theorem 1 with $d = 3$ applies to the special cases of bar-joint systems and body-pin systems essentially equivalent to body-bar systems. We combine these characterizations into the theorem below, by using a general sparsity constant $l_C$ which depends on the particular class $C$ of D-reps under consideration. For the $d = 2$, bar-joint class $C$, the constant $l_C = 3$.

**Theorem 1** (Graph characterization of Minimal Rigidity). *A constraint graph $G = (V, E)$ from a nice class $C$ is generically minimally rigid in d dimensions if and only if for a small constant $l_C$ depending only on the class $C$ of D-rep,*

$$|E| = d|V| - l_C, \quad and \quad |E'| \leq d|V'| - l_C$$

*for every subgraph $G' = (V', E')$.*

Nice minimally rigid constraint graphs $G$ therefore have an average degree close to $2d$.

### 2.4. Computational Efficiency

The above discussion shows that many properties of D-reps are graph properties of the underlying constraint graph $G$. For example, using Theorem 1, for many nice classes $C$, there are efficient, essentially greedy algorithms for determining if the graph is minimally rigid, determine rigidity percolation etc. Most are based on special cases of network flow called pebble games [16, 25, 20]. Furthermore, there are efficient algorithms for recursively decomposing such constraint graphs into maximal proper subgraphs that are also minimally rigid, towards efficient realization of the corresponding D-reps [2, 16, 15].

Even physical properties that are not combinatorially characterizable, i.e, depend on the geometry of the D-rephave highly efficient computations as will be discussed in Section 4.4.

## 3. Design Operations

We propose a sequence of operations for tailoring microstructure at several scales to a given macroscale shape and to physics requirements. We keep the setup simple by assuming that we can

1. build the entire model as images of subdomains $\Omega_i$ that are cubes, prisms or tetrahedra (see Fig. 2a). These three types of subdomains are each mapped by some regular function $f_i : \Omega_i \to \mathbb{R}^3$ to *subregions* in physical space so as to conform to the macro-shape surface and possibly respond to coarse field lines in the interior.

Alternatively, existing solids can be partitioned into subregions parameterized over subdomains [29, 33]. Each subdomain type has natural families of iso-parameter planes, three for the hexahedral and four for the prism and tetrahedron. Since the $f_i$ are regular,

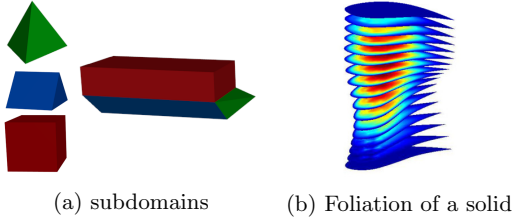

(a) subdomains   (b) Foliation of a solid

Figure 2: Preparation steps for applying D-reps.

2. each family of iso-parameter planes induces a *foliation* of the subregion such that the image of an iso-parameter plane is a possibly curved foliation surface and macro-shape surfaces are images of subdomain faces (see Fig. 2b).

To foliate more generally, e.g. following a trivariate field, see [4]. E.g. [30] shows how to compute fields on foliated irregular shape domains. The focus of this paper is on steps 3 and 4.

3. For each iso-parameter plane select and generate a regular bivariate D-rep micro-structure (see Section 4). Introduce non-uniform connectivity or refine until the foliation surface matches the material or physics requirements (see Fig. 3, Section 4.2 and Section 4.3).

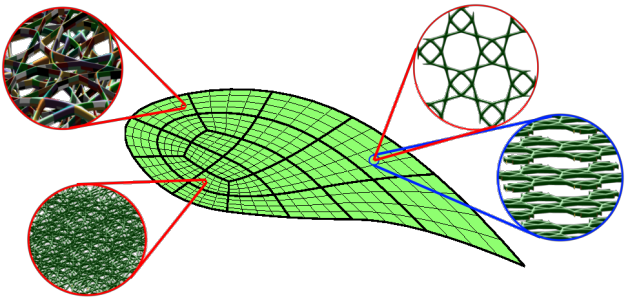

Figure 3: Choice of bivariate micro-structures. Enlargements: *red (right)* shows regular micro-structure, *red (left)* show irregular micro-structure viewed from top, *blue* shows the connected foliation stack.

4. Re-assemble adjacent subregions to form the macro shape, possibly adjusting the micro-structure near interfaces (see Section 4.5 and Section 4.6).

Since steps 3 and 4 are carried out in the domain of a continuous map $f$ that that is the union of the maps $f_i$, properties that depend only on the constraint graph are preserved under $f$.

## 4. The CoST Design Space

We now exhibit rich design space all of whose members are *minimally rigid*, i.e. neither over-constrained (under self-stress) nor under-constrained (mechanisms) and admit multiple types of physical constraints and material realizations.

**Definition 2** (CoST). *A k-regular graph is graph where every vertex has k neighbors. The graph $K_n$ is the complete graph on n vertices, i.e., it contains all edges. A bivariate* Corner Sharing Triangle *(CoST) graph G is a 4-regular graph such that every vertex is part of exactly two edge-disjoint $K_3$ (triangle) graphs. A trivariate* Corner Sharing Tetrahedron *(CoST) graph G is a 6-regular graph such that every vertex of G is part of exactly two edge-disjoint (tetrahedral) $K_4$ graphs.*

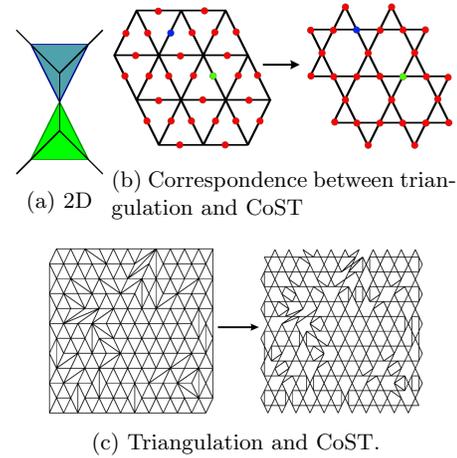

(a) 2D   (b) Correspondence between triangulation and CoST

(c) Triangulation and CoST.

Figure 4: Every 3-regular vertex spawns a corner-sharing triangle. (a) zoomed in detail, (b) procedural bijection on a larger triangulation.

Fig. 4(b) illustrates for planar graphs how connecting the mid-edges emanating from a vertex into a simplex yields a bijection from a 3-regular graph to a bivariate CoST.

**Observation 3** (Bijection from CoSTs To Regular Graphs). *There is a bijection between bivariate CoSTs and 3-regular graphs. There is a bijection between trivariate CoSTs and 4-regular graphs.*

Since we will build trivariate CoSTs from bivariate ones, and since *planar* (noncrossing) bivariate CoSTs can be particularly well understood, we first characterize these. Planar CoST graphs are 'witnessed' by set $W$ of edge-disjoint triangular facets such that every internal vertex is shared by exactly two triangular facets in $W$ and every boundary vertex has degree 2.

**Theorem 4** (Planar Bivariate CoSTgraphs are rich).

- *The bijection in Observation 3 - when restricted to planar 3-regular graphs where every boundary vertex has degree at least 2 - maps to planar bivariate CoST graphs.*

- *There is a bijection between planar bivariate CoST graphs and planar triangulation graphs.*

Finite planar bivariate CoSTs have boundary vertices of degree 2 that need to be pinned or connected ('stiffened') to make them minimally rigid. Section 4.1 discusses multiple boundary constraints that guarantee minimal rigidity.

We now generate trivariate from bivariate CoSTs (see Fig. 6).

**Definition 5** (Trivariate CoST). *A bivariate planar CoST is 2-colored if no two vertex-sharing triangles have the same color. Consider a sequence $C_1, C_2 \ldots$ of 2-colored bivariate planar CoSTs, each with g green and b blue triangles. For i even, between every pair of a blue triangle in $C_i$ and a blue triangle in $C_{i+1}$ introduce a new vertex and connect it to all six triangle vertices (see Fig. 6). For i odd introduce and connect such a vertex only for the green pairs. The resulting trivariate graph is a* foliated trivariate CoST graph.

There exist other 4-regular planar and 6-regular foliated graphs, for example the quadrilateral grid on a torus. But these lack the richness of CoST graphs and can be converted , in multiple ways, to CoSTs as shown in Fig. 3.

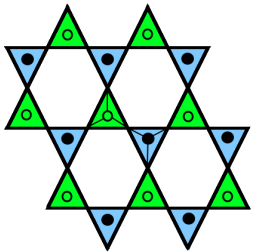

Figure 5: Construction of trivariate CoSTs of Definition 5. Circled point are below, solid points above the plane of the bivariate CoST.

### 4.1. CoSTs from Kagome Seeds

We introduce geometry starting with uniform *seed* sequences of unit-length, balanced, bar-joint CoSTs. From these seeds, we generate more general CoSTs. We choose *bar-joint* D-reps because they have the highest versatility to represent D-rep types (body-bar, body-pin, tensegrity) and physics (mechanical, electrical, thermal). A CoST is called *unit-length* if all bar lengths within a subregion are approximately equal or chosen from a finite set. This is natural for both designed meta-materials and naturally occurring materials (Silicon and Carbon based, sticky sphere colloids, jammed sphere composites). A CoST is *balanced* if it is realizable as a 2D (resp. 3D) tensegrity D-rep with only struts or only ties (e.g. with material realization as packed spheres). This means that in order to resolve stresses at a joint, any line (resp. plane) through the joint should have at least one bar vector positioned on either side. The most natural such CoSTs turn out to be a class of CoSTs known as bivariate Kagome lattices and their trivariate generalization (Fig. 6).

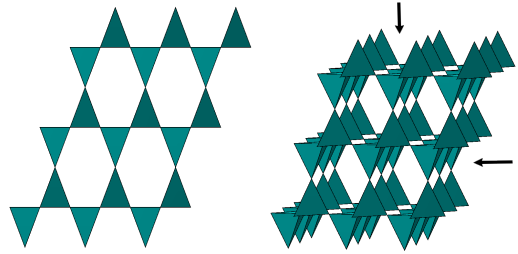

Figure 6: Regular bivariate and trivariate CoSTs: bivariate Kagome and foliated trivariate Kagome. There are four direction from which the trivariate Kagome looks like the bivariate Kagome.

**Definition 6.** *A bivariate Kagome CoST forms a hex-tri partition of the plane (Fig. 6). The* trivariate Kagome CoST *is a foliated trivariate CoST obtained from any one of its four sequences of parallel bivariate Kagome CoSTs as in Definition 5.*

As truss and wire woven structures, Kagome lattices have been studied extensively and their physical properties have been characterized as superior to Octet and other truss structures Lee and Kang [26], Hyun et al. [18].

**Observation 7.** *Alternatively, the trivariate Kagome is obtained as the intersection lattice of four sequences of parallel bivariate Kagome D-reps.*

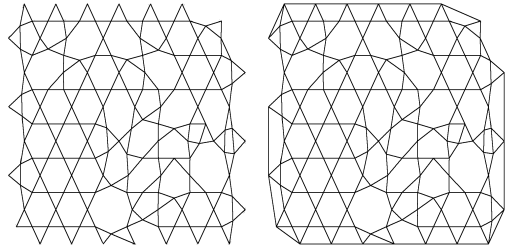

Figure 7: Stiffening bivariate CoSTs

Along boundaries, bivariate CoST graphs have vertices of degree 2 and 3. Hence they are not minimally rigid.

However, the following algorithm(s) adapted and extended from [46] provides three ways of correctly *stiffening* the boundary of a planar, bivariate CoSTs so that they are minimally rigid.

**Algorithm 4.1.** [Stiffening]
*Input* A finite piece of a bivariate CoST $F$ with graph $G$.
*Output* A graph $G'$ that is minimally rigid for a class $C$ (with constant $l_C$ of Theorem 1).
*Algorithm* for constructing $G'$.
Label the boundary vertices of $G$ in sequence as $b_1, \ldots, b_m = b_1$ and set $B := \{b_1, \ldots, b_m\}$. Pick $l_C$ distinct vertices from $B$ to form the subset $A := \{a_1, \ldots, a_{l_C}\} \subset B$.

i. For $b_i \in B \setminus A$ add an edge from $b_i$ to $b_{i+1}$ and $b_{i-1}$; Connect each $a_j \in C$ to its predecessor.

ii. for $b_i \in B \setminus A$ fix the 2 coordinates of $b_i$.

iii. for $b_i \in B \setminus A$ constrain the 2 coordinates of $v_i$ to lie on a fixed line (slider) $l_i$ on the plane. □

For trivariate CoSTs, boundary stiffening is more involved since there is no general theorem characterizing generic rigidity of bar-joint frameworks in 3 dimensions. However, since trivariate CoSTs underlying periodic Kagome seeds are a special class of 6-regular graphs, namely body-pin graphs, with exactly 2 bodies sharing a pin, no two bodies sharing more than 1 pin, and no proper subgraph that is generically rigid, a rule analogous to (i) in the above algorithm can be used for stiffening. See Figure 8.

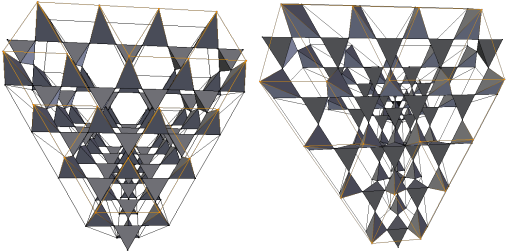

Figure 8: Stiffening a boundary (for a trivariate CoST) by judiciously adding edges to the boundary.

*4.2. CoST Refinement and Hierarchy*

We consider the refinement operation separately for bivariate and trivariate CoSTs; and separately for the graph-theoretic and for the geometric representations.

We generate multiple bivariate CoST hierarchies with self-similar structure. Akin to interpolatory subdivision, we retain all vertices and so preserve the vertex degree and edge incidence obtained from splitting incident edges at the once-coarser level. This *constructive refinement* yields a recursive block in the rigidity, stiffness, Laplacian, and related matrices (cf. Section 2) and so permits efficient computation of relevant physical properties at increasing levels of the hierarchy.

We assume that all boundary vertices have degree two. Since the operations described below preserve the CoST properties, their boundary can be stiffened as in Section 4.1 to ensure minimal rigidity.

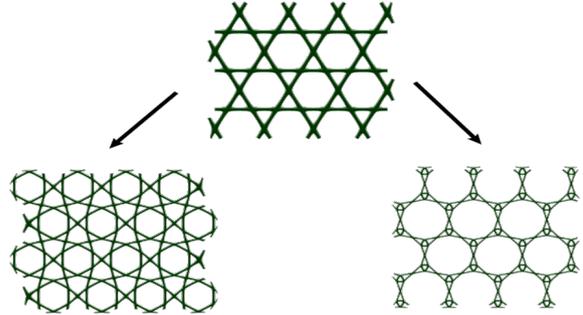

Figure 9: Rule $R_1$ (*left*) vs $R_0$ (*right*)

**Definition 8** (Bivariate Refinement Rules $R_k$). *For each facet being refined, split each edge $(u, v)$ of the witness set $W$ of the planar, bivariate, CoST into two edges $(u, z)$ and $(z, v)$ by adding a new vertex $z$. Use the clockwise ordering of edges in the facet to add new edges to connect the newly added vertices in a cycle. The rule is denoted $R_0$ when applied to a triangle in $W$ and $R_1$ otherwise.*

The rules create two new triangles per old vertex. These triangles form a new witness set $W$.

**Lemma 9.** *The bivariate CoST refinement rule preserves the CoST property.*

The following observation applies to bar-joint, body-pin and tensegrity CoSTs.

**Observation 10.** *Let $F$ be a bivariate, unit-distance and balanced CoST. Applying Rule $R_0$ and positioning new vertices at the midpoint of the old bar ensures that the refined CoST is unit-distance. (Rule $R_1$ destroys unit-distance.) Both rules require moving the new point to the centroid of its neighbors to ensure that the CoST is balancedwhich also destroys unit-distance.*

**Definition 11** (Trivariate CoST Refinement). *Apply Rule $R_0$ adding midpoints to the edges of the bivariate CoSTs of a trivariate CoST. Restore balance by moving each new point to the centroid of its neighbors.*

Alternatively, the foliated CoST can be assembled after the desired level of refinement has been performed on all bivariate CoSTs.

Trivariate refinement creates four new tetrahedra for each old tetrahedron and creates two new layers between any two parallel bivariate CoST layers. The refinement preserves unit distance and foliation structure.

*Mass Computation and Design.* CoSTs represent many possible materials. Measuring change in mass under refinement is a simple computation regardless whether the mass is at vertices, edges or faces of the bivariate or trivariate D-rep.

**Observation 12.** *Every vertex of the CoST graph belongs to exactly two simplices of the witness set W; every edge and face belong to exactly one.*

Therefore the perimeter/area/volume measure of the simplices in $W$ provides an accurate measure of relative or proportional change in mass during refinement. Conversely, we can choose the refinement rule $R_0$ or $R_1$ at each step of the constructive refinement to achieve the desired change of mass.

### 4.3. Fully Exploring the Space of CoSTs and Modeling Random Irregularities

The results in this section serve two goals. The first is to deterministically alter the current CoST graph and possibly its geometry – while maintaining CoST and other relevant local geometric constraints to ensure that geometric constraints (point incidence, unit-distance, and balance) have a realization. The second is to generate a CoST according to a distribution of naturally occurring materials or to tailor random matrix composites to macroscale shape and physics.

We leverage a classical result for planar triangulation graphs. Given two vertex-labeled planar triangulation graphs $A$ and $B$ on $n$ vertices, there exists a path of planar triangulation graphs $A, A_1, \ldots, B$ where $A_{i+1} = A_i \cup (u,v) \setminus (w,x)$ and triangles $(u,v,w)$ and $(u,v,x)$ appear as facets in some plane embedding of $A_i$. Switching the edge $(u,v)$ with the edge $(w,x)$ is called a *diagonal flip*. Since there is a bijection from planar triangulation graphs to bivariate planar CoSTs, we use the same term to refer to the corresponding operation on CoSTs of all its bivariate foliations (see Section 4). By Definition 5 we can extend this result to corner sharing tetrahedra as a *trivariate diagonal flip* (See Fig. 11c, Fig. 11d).

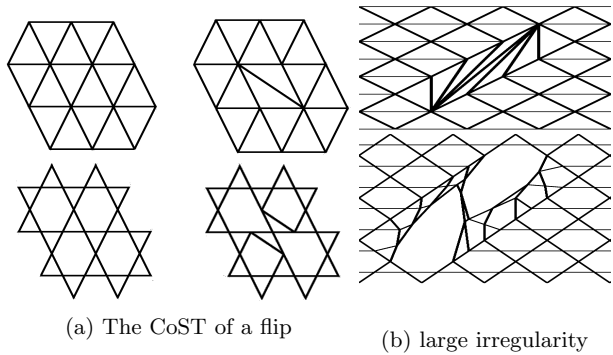

(a) The CoST of a flip    (b) large irregularity

Figure 10: (a) Effect of one diagonal flip on the three-direction triangulation (*top*) and its corresponding CoST (*bottom*). (b) A large irregularity in (*top*) the triangulation and (*bottom*) the corresponding trihex CoST graph (and restoring balance). Note that the graph is still minimally rigid.

**Observation 13.** *We can explore all CoST graphs using diagonal flips.*

Given a fixed point set, we explore all bivariate geometric CoSTs, by exploring all triangulations reached by flips on adjacent triangles that form strictly convex quadrilaterals [34]. This automatically enforces balance. Since there is always a triangulation that contains any given line segment connecting any given pair of points in the point set, even triangulations with skinny triangles can be modeled this way (Fig. 10). Uneven connectivity can for example introduce large cylindrical holes in the regular trivariate Kagome D-rep as well as structures that no longer obey the foliation directions Fig. 11b.

Diagonal flips underlie so-called Stone-Wales defects in many types of layered crystals based on the regular triangular lattices. Since any non-edge in a triangular lattice can be represented by relatively prime coordinates $(p,q)$, a generalized Euclid's algorithm and Farey sequences can be used to generate a sequence of neighboring diagonal flips that result in flipping an arbitrarily long non-edge into an edge. This gives provable bounds on the number of defects that lead to non-edges or connected channels of the required length (see Fig. 10). Combined with stiffness analysis of Section 4.4) this gives a highly efficient way to obtain rough estimates while modeling fracture.

The diagonal flip also provides an efficient and principled way of modeling *random micro-structure with a predictable output distribution*. For example, a Poisson process can determine independent locations for the diagonal flip, or a Markov process can increase the probability of generating diagonal flips closer to previous flips in order to model fracture. Seeds followed by $n$ steps of the two refinement rules of Section 4.2 can so model $2^n$ samples of random CoST micro-structure.

### 4.4. Static Multiphysics: Stiffness, Laplacian, Resistance, Mass

The GCS of a constraint graph $G$ is a set of polynomial equations $F = \{f_1, \ldots, f_m\}$ and variables $X = \{x_1, \ldots, x_n\}$. The *rigidity matrix* $R_G(p)$ is the Jacobian of this system with respect to $X$ evaluated at instantiations $p(x_i)$ for variable $x_i$ is

$$R_G(p) \in \mathbb{R}^{m \times n} : (i,j) \to \frac{\partial f_i}{\partial x_j}(p).$$

Since constraint systems are typically quadratic, $R_G(p)$ is often referred to as *linearization* of the polynomial system. Assuming generic $p$, row independence of $R_G(p)$ corresponds to local independence of the algebraic constraints of the original D-rep [1], and rigidity of the D-rep $(G,p)$ is equivalent to $R_G(p)$ having maximum possible rank (over all D-reps $(G,p)$). Since maximum rank over $p$ is attained at all generic $p$ it and the associated properties of independence and rigidity of the original D-rep depend only on the constraint graph $G$.

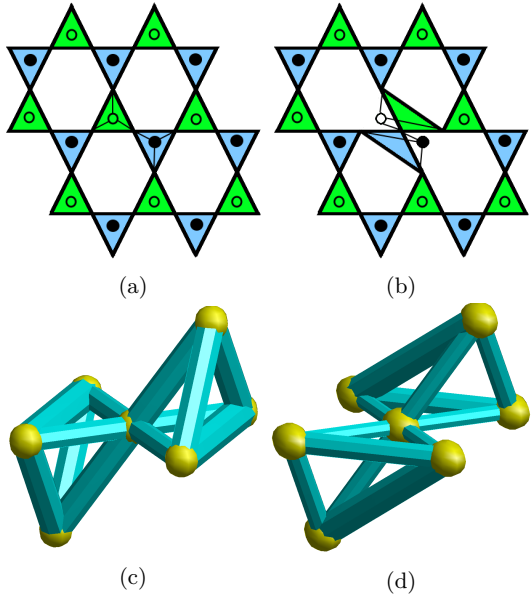

(a) (b)
(c) (d)

Figure 11: Irregularity in a trivariate CoST. (*top*) top-down view, green tetrahedra face down and connect to empty circles, blue tetrahedra connect up to solid circles. (*bottom*) Effect of a flip in a trivariate foliated CoST.

Geometrically, any (infinitesimal) vector in the right nullspace of $R_G(p)$ represents an (infinitesimal) change to each primitive such that the resulting D-rep still satisfies all of the constraints. The left nullspace corresponds to an *equilibrium self-stress* of the system [40, 7]. In the case of bar-joint or tensegrity D-reps, these stress vectors correspond to length-normalized forces on the bars, struts or ties that cancel out at the joints. Moreover, the well-known stiffness matrix (when the bars/struts/ties are replaced by springs) has the same null space as the rigidity matrix. The stiffness matrix is therefore essentially a weighted graph Laplacian of the constraint graph $G$ underlying the D-rep. The Laplacian also yields the effective electrical resistance and analogously thermal resistance of appropriate physical realization of the D-rep. Further related analogs include so-called topological insulator dielectric properties as well as photonic and acoustic properties related to $G$ [13, 31, 32].

Mass generally enters the conversion of abstract D-rep properties into analogous physical material properties. For example, to achieve a physical stress $s$ on a material realization of an abstract bar with stress $t$ and length $l$ in a bar-joint D-rep, the cross-sectional area $A$ of the material bar must be $lt/s$. The desired strain, effective Young's modulus and density of the bar material follows from setting its mass.

The rigidity, stiffness and Laplacian matrices of a D-rep, and their null spaces defining stresses, flexes etc. provide an efficient way of computing effective multi-physical properties without finite elements.

### 4.5. Joining CoSTs

When joining sub-regions the combined D-rep should again be a CoST, i.e. be minimally rigid, unit-distance and balanced. We focus on matched foliations - joining unmatched

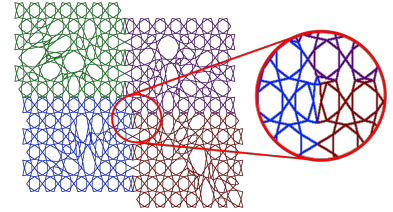

Figure 12: Joining bivariate CoSTs

foliations may require modified stiffening of the denser subregion.

The problem then reduces to joining bivariate CoSTs that have the same number of degree 2 boundary vertices after removal of the additional stiffening edges of Algorithm 4.1. Then degree 2 vertices, one from each abutting CoST, are pairwise merged (Fig. 12) into a single vertex to restore 4-regularity and minimal rigidity of the combined CoST graph. The geometric realizations require only a recomputation in a small neighborhood of the merged vertices.

### 4.6. Re-realization

Joining unmatched CoSTs, changing of connectivity, constructive refinement and distortion by the maps $f_i$ may require local bivariate re-computation of the node placement to avoid vastly uneven edge lengths. In general, solving 2D distance constraints is computationally demanding process equivalent to solving a quadratic polynomial system, an NP-complete challenge. Fortunately, subsystems are small due to locality and have a structure that makes them tractable by recent, nearly linear time algorithms, e.g. CayMos [44, 50, 41]. This class of algorithms reduce the computation of realizations with known orientations, to searching the space of parameterized ruler-and-compass constructions.

## 5. Combining with Continuous Representations

Some operations in CAD systems require a boundary representation (B-rep such as NURBS), e.g. for display. Analysis of physics in 3-space is convenient when based on trivariate fields, e.g. a homogenized implicit representation of micro-structure. These representations associated with the CoST graph then admit higher-order iso-parametric or iso-geometric analysis.

A *parametric* interpretation simply thickens the edges into beams (see Fig. 13(a)). This representation can also be leveraged for generating additive layer deposition planes as illustrated in Fig. 13 (b). While a beam representation suffices for initial display, we implemented a simple, local algorithm that fashions each half of a beam from four $C^1$-joined bi-quadratic patches. These bi-quadratic cylinder-like pieces then join to enclose the node and join $C^1$ with the pipe piece of the neighbor node. The construction is local, linear in the location of the nodes and offers parameters to locally thicken or thin the beams. Fig. 14a illustrates the continuous gradation of a bivariate CoST surrogate. The arrow in the figure points to a modeled reinforcement of a particular pair of edges.

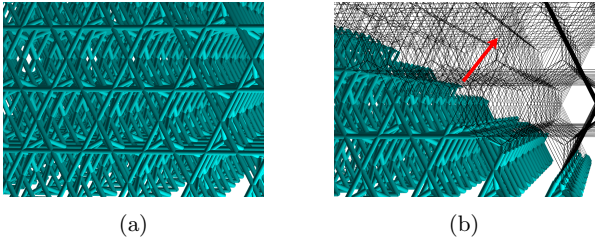

(a)          (b)

Figure 13: Regular Kagome micro-structure. (b) Cut along a (n additive layer deposition) plane orthogonal to the red arrow.

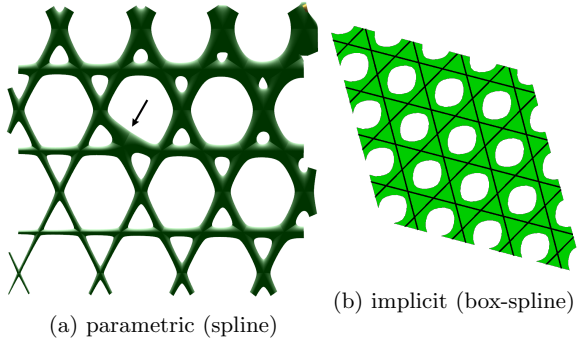

(a) parametric (spline)      (b) implicit (box-spline)

Figure 14: (a) Gradation and local adjustment of the bi-quadratic parameterized surface representation associated with a geometric realizations of the constraint graph. (b) 3-direction box-spline level set.

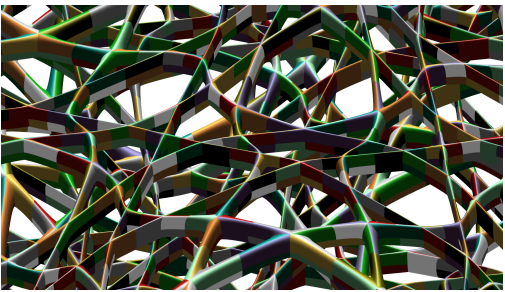

Figure 15: Parametric spline associated with bivariate CoST layers from Fig. 3. Each spline patch is shown in a different color.

Since the surface is polynomial, it is straightforward, via the divergence theorem, to compute the volume contributed by each cylinder-like piece. By symbolic computation, the contribution can be expressed as an inner product (less than 100 add or multiply per graph edge) of a precomputed vector representing the known structure and determinants of the node and its direct neighbors weighted by the parameters.

A smooth volumetric representation can be associated with a geometric realizations of a regular graph by interpreting the nodes and their values (for example mass) as box-spline coefficients [8]. In two variables, the three directions of the trihex CoST are naturally associated with the convolution directions of the 3-direction box spline of the hat function [8]. Double convolution in each of the three directions yields a $C^2$ box spline. The bivariate field is of total degree 4. Its zero level set when setting the node values of the trihex CoST to 1 and the empty grid locations to -1 is displayed in Fig. 14b. The generalization of this box spline to irregular meshes is known as Loop subdivision [27].

In three variables, the four planes of the Kagome CoST suggest interpreting the nodes as control points of a 4-direction box spline (see e.g. [23]). Double convolution in the directions yields a smooth volumetric field whose pieces are of total degree 5. Fig. 16a shows in yellow the zero level set when setting the node values to 1 and the empty grid locations to -1. Fig. 16b shows the Kagome CoST as blue tetrahedra superimposed on the level set, filling in the voids with value greater than zero. Specifically when the underlying constraint graph has uniform distances and Kagome connectivity, this models Tridymite, a quartz-like material that owes its strength to its microstructure. Fig. 17 illustrates how the Kagome microstructure in a subdomain is reinterpreted as a box-spline volumetric function (with yellow zero set) and mapped to an octant of a sphere.

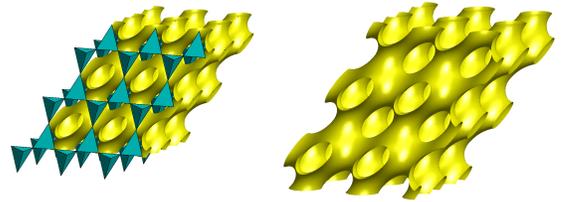

Figure 16: Zero level set (yellow) of the trivariate 4-direction box-spline initialized as values 1 at the nodes of the Kagome CoST (blue tetrahedra).

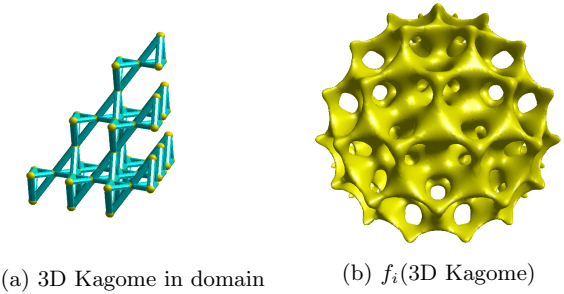

(a) 3D Kagome in domain     (b) $f_i$(3D Kagome)

Figure 17: Tetrahedral domain mapped to an octant of a sphere

To switch computation from the CoST to continuum representation and volumetric finite elements (based on the implicit representation) established via homogenization. The traditional engineering challenge when modeling naturally occurring materials is to choose the representative volume elements (RVEs) sufficiently large so that the RVE correctly models effective (averaged) properties [35]. For CoSTs, the process is reverse: we refine until a fixed sub-region can be treated as RVEs. However if material constraints, for example, particulate sizes in matrix composites, do not permit further refinement then the subregion cannot be homogenized, but must be treated explicitly as a CoST.

## 6. Conclusion

We introduced tools for efficient design and analysis of micro-structure via corner sharing triangles and tetrahedra (CoSTs), a family of discrete geometric constraint representations. CoSTs allow modeling micro-structure via a number of simple operations: graph manipulation, constructive, hierarchical refinement, generation of irregularities according to distributions, and combining with continuous representations.

CoSTs form a rich space: bivariate CoSTs are as rich as triangulations. We therefore suggest initializing regular CoSTs as Kagome lattices. CoSTs can represent many types of local constraints that imply global properties of multi-scale structures for efficient and direct design and analysis and minimal rigidity in particular. A CoST's rigidity, stiffness and Laplacian matrices $R_G$ and its left and right null spaces define stresses and flexes without finite elements. And CoSTs admit a self-similar or hierarchical structure that retains information from previous levels. In particular the corresponding recursive block structure is inherited by the above matrices resulting in significant computational efficiency [2]. Since a sequence of flips can model the emergence of material defects, we therefore see potential for *efficient* fracture analysis. We also expect that the discrete representation, together with control over the distribution of irregularities in CoSTs will allow an efficient computation of Representative Volume Elements (RVEs) for homogenized representation of bulk properties. This would facilitate combining the CoST-scale microstructure analysis with the Finite Element Method at a larger scale of bulk material.